\documentclass[11pt]{article}
\usepackage{graphicx}
\usepackage{amsmath}

\title{ ESTIMATION OF RADIATIVE CORRECTIONS TO THE PROCESS OF MUON-ELECTRON CONVERSION }

\author{\textbf{Rashid M. Djilkibaev$^{1}$ \thanks{Permanent address:
     Institute for Nuclear Research, 60-th Oct. pr. 7a,
     Moscow 117312, Russia}\ ,
     Rostislav V. Konoplich$^{1,2}$}\\
   \normalsize$^{1}$Department of Physics, New York University,
   New York, NY 10003\\
   \normalsize$^{2}$Manhattan College, Riverdale, New York, NY, 10471}

\begin{document}

\maketitle

\begin{abstract}

In detection of electrons from 
$\mu \to e $ conversion process
the monochromatic electron spectrum is transformed
due to a photon emission and 
fluctuations of energy loss in a  target.
The selection criterion
of $\mu \to e (\gamma)$ conversion events is an electron momentum above
the threshold momentum of 103.5 MeV/c , which corresponds to the maximum 
energy loss of 1.5 MeV. 
Radiative corrections including a virtual photon correction and 
soft photon emission 
below 1.5 MeV lead to a reduction by about 10$\%$ in
the probability of $\mu \to e $ conversion process calculated 
without radiative corrections.

The soft photons emission below 1.5 MeV
contributes to a change of electron spectrum from monoenergetic one 
at 105 MeV to a spectrum with a low energy tail for the process of 
$\mu \to e (\gamma)$ conversion. 
However the effect of smearing of the initial momentum distribution
due to the soft photon emission is small in
comparison with a smearing due to energy loss fluctuations in a target. 
The average energy of soft photons emitted below 1.5 MeV is found to be 40 keV. 
The soft photon approximation is a good description for  
$\mu \to e (\gamma)$ conversion process with photons emitted
below 1.5 MeV.

\end{abstract}

\newpage

\section*{Introduction}

In this work we study radiative corrections to the process of   
$\mu \to e $ conversion, i.e. we consider the process of 
$\mu \to e (\gamma)$ conversion including the elastic process of 
$\mu \to e $ conversion and the process of $\mu \to e$ conversion
with a photon emission below a certain energy $\omega_{max}$.
The $\mu \to e $ conversion is the process of a transformation of a muon 
into an electron in the presence of a nucleus. 
The clear signature of $\mu \to e $ conversion is an appearance in the 
final state of a monochromatic electron of the energy

\begin{equation}
 E_{max} = E_{\mu} - \frac{E_{\mu}^{2}}{2 \cdot M_{A}},
 ~~~~~~~~~~E_{\mu} = m_{\mu} \cdot(1 - (\alpha \cdot Z)^{2}/2) 
\label{Emax} 
\end{equation}

where $M_{A}$ is a nuclear mass, $m_{\mu}$ is a muon mass,
$\alpha$ is the fine structure constant, Z is a nuclear charge.
In particular, $E_{max}$ is 104.963 MeV for aluminum.

In an experimental setup a monochromatic spectrum of
$\mu \to e $ conversion electrons
at 105 MeV is transformed to a distribution with a low energy tail.
This transformation in the electron spectrum can be either due to 
fluctuations of energy loss of electrons in a target
 or because of a photon emission 
in the process of $\mu \to e (\gamma)$ conversion. The selection criterion
of $\mu \to e (\gamma)$ conversion events is an electron momentum above
the threshold momentum of 103.5 MeV/c , which corresponds to the maximum 
energy loss of 1.5 MeV. 

It is important to note that in real experiments a pure elastic
process is not physically observable since due to the presence
of charged particles in the initial, intermediate or final 
states the process can be accompanied by
an emission of a photon.  Therefore the measured probability 
of the process of $\mu \to e (\gamma)$ conversion 
is a sum of probabilities of the pure 
elastic process corrected
by virtual photon exchange and of a real soft photon emission
below a maximum energy $\omega_{max}$, which in our case is 1.5 MeV.
In fact if the emitted real soft
photon is not detected, one can not distinguish a bremsstrahlung
process from the elastic one.

Because the energy release in the process of 
$\mu \to e (\gamma)$ conversion is about $\Delta E = 105 ~MeV$ one could 
expect that the virtual photon corrections and the soft photon 
emission
can be enhanced by a large logarithmic factor $ln(m_{\mu}/m_e)$ 
where $m_e$ is an electron mass. 
Radiative corrections lead to a change in
the probability of $\mu \to e $ conversion process calculated 
without radiative corrections.

The real soft photons emission 
contributes to a change of electron spectrum from monoenergetic one 
at 105 MeV to a spectrum with a low energy tail. 
However this effect of smearing of the initial momentum distribution
due to the soft photon emission 
should be compared with a smearing due to energy loss 
fluctuations in a  target. 

The detailed analysis of hard photon emission requires knowledge 
of specific particle physics models. In order to avoid
complications in analysis of high energy photon spectrum
and to understand qualitatively a hard photon emission 
the radiative pion decay is considered.
By analyzing this decay which possesses the same kinematic 
properties and a comparable energy scale as $\mu \to e$ conversion 
the range of application of the soft photon approximation 
can be found.

It is worth to note that
the hard photon emission with photons produced at energy 
above 1.5 MeV is not important for simulation of 
$\mu \to e (\gamma)$ conversion process. 
Hard photons carry away a 
significant part of muon energy thus reducing an electron energy 
but due to the selection criterion 103.5 MeV/c on 
the threshold momentum such electrons
are rejected and such events do not contribute to useful events.

\section*{Virtual photon corrections}

\begin{figure}[htb!]
  \centering
  \includegraphics[width=0.7\textwidth]{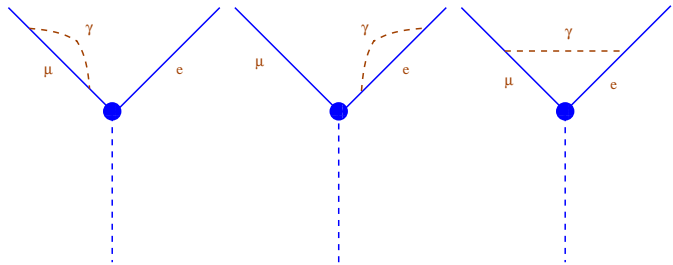}
  \caption{
Virtual photon corrections to lepton vertex and ingoing and outgoing 
leptons in 
the process of muon-electron conversion. 
 }
\label{fig:diag_virt}
\end{figure}

Let us consider $\mu \to e$ conversion process in which a virtual photon
emission is involved. 
The Feynman diagrams describing virtual photon emission ~\cite{landau}  are presented
in Figure ~\ref{fig:diag_virt}.  

It can be shown that an integral describing the radiative correction
to the lepton vertex is infrared divergent. This integral can be
artificially well-defined with an introduction of an infinite 
small mass $\lambda$ of a photon. With this modification the leading term
in leading logarithm approximation is given by

\begin{equation}
\Gamma_{virt}=\Gamma_{el} \cdot \left [1-\frac{2\alpha}{\pi}ln\frac {m_{\mu}}{m_e}ln\frac{m_{\mu}}{\lambda} \right ]
\label{virt}
\end{equation} 

where $\Gamma_{el}$ is the $\mu \to e$ conversion rate calculated
without radiative corrections.
\section*{Soft photon emission}

\begin{figure}[htb!]
  \centering
  \includegraphics[width=0.7\textwidth]{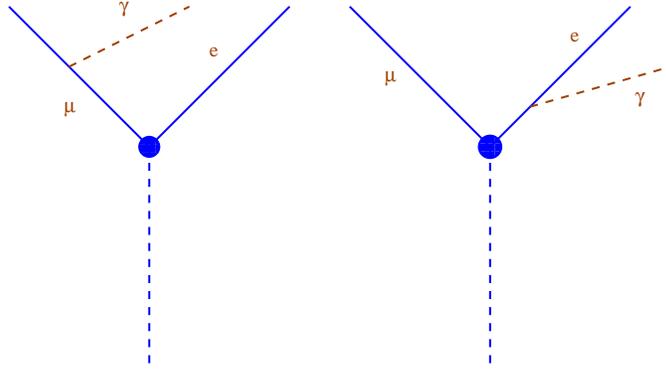}
  \caption{
Inner bremsstrahlung in 
the process of muon-electron conversion. 
 }
\label{fig:diag_soft}
\end{figure}

Feynman diagrams describing the process of inner bremsstrahlung are presented
in Figure ~\ref{fig:diag_soft}.  

Let's consider the case of emission of low energy photons. In this limit we
assume that a photon energy is much less than energies of particles taking part 
in muon-electron conversion and momentum transfered to a nucleus. On the other 
hand we assume the validity of the perturbation theory, i.e. the smallness of
a parameter $\alpha \cdot ln(m_{\mu}/m_e) << 1$ which provides a suppression  
of multi-photon emission.

Following a standard technique (see e.g.~\cite{landau} )
in the limit of photon low energies the photon emission can be taken into 
account by adding an external photon line to a diagram of the elastic process.
In this case the most important are diagrams with an emission from external
charged particles lines, in all other parts of diagrams one can neglect
changes in momenta due to the soft photon emission.
Therefore Dirac spinors of initial muon and outgoing electron
are modified in the following way

\begin{center}
$u_{\mu} \to e \cdot \frac{\hat k - \hat q + m_{\mu}}{(k - q)^2 - m_{\mu}^2} \cdot \hat \varepsilon ^* u_{\mu} 
\approx -e \cdot \frac{\hat k + m_{\mu}}{2(kq)} \cdot \hat \varepsilon ^* u_{\mu} ~~,$ 
\end{center}

\begin{center}
$\bar u_{e} \to  e \cdot \bar u_{e} \hat \varepsilon ^* \frac{\hat k_1 + \hat q + m_{e}}{(k_1 + q)^2 - m_{e}^2} 
\approx e \cdot \bar u_{e} \hat \varepsilon ^* \frac{\hat k_1 + m_{e}}{2(k_1q)} ~~,$
\end{center}

where k, $k_1$, q are the four-vectors of muon, electron and photon, respectively,
$\varepsilon$ is the polarization of a photon.
A photon emission from a non-relativistic nucleus is neglected.

By using Dirac equations these relations can be simplified 

\begin{center}
$u_{\mu} \to -e \cdot \frac{(k\varepsilon ^*)}{(kq)}  u_{\mu} ~~,$ 
\end{center}

\begin{center}
$\bar u_{e} \to e \cdot \bar u_e \cdot \frac{(k_1\varepsilon ^*)}{(k_1q)} ~~.$ 
\end{center}

As a result amplitude M of muon-electron conversion with the soft 
photon emission can be factorized according to 

\begin{equation}
M = M_{el} \cdot e \left [ \frac{(k_1\varepsilon ^*)}{(k_1q)} - \frac{(k\varepsilon ^*)}{(kq)} \right ] ~,
\label{matr}
\end{equation}

where $M_{el}$ is the amplitude of the same process of $\mu \to e $ conversion 
without photon emission. This is well-know general result which is process
independent. 

Summing up over the photon polarization and by applying standard technique of 
calculation of process probability the differential probability to emit a single
soft photon with momentum q in the process of $\mu \to e$ conversion is given by

\begin{equation}
d^9 \Gamma_{soft} = d^6 \Gamma_{el} \cdot 4 \pi \alpha \left [ \frac{2(k_1k)}{(k_1q)(kq)} - 
\frac{(m_e^2)}{(k_1q)^2} - \frac{(m_{\mu}^2)}{(kq)^2} \right ] \frac{d^3q}{(2\pi)^3 2\omega} ~ .
\label{prob1}
\end{equation}

Integration of Eq.(\ref{prob1}) over an electron momentum within kinematic
limits gives the spectrum of 
soft photons in the process of muon-electron conversion

\begin{equation}
\frac{1}{\Gamma_{el}} \frac{d\Gamma_{soft}}{d\omega} = \frac{2\alpha}{\pi} \cdot 
ln(\frac{m_{\mu}}{m_e}) \cdot \frac{1}{\omega} ~.
\label{prob2} 
\end{equation} 

Note that the same expression for the spectrum of soft photon emission can be 
obtain by applying the classical consideration ~\cite{jackson}
instead of Feynman diagram technique.

The limit $\omega \to 0$ in Eq.(\ref{prob2}) represents a logarithmic divergence meaning that the total 
probability of emitting a single very soft photon by a charge particle is infinite,
this is well-known infrared catastrophe. By introducing a small photon mass 
$\lambda$ the integral over $\omega$ becomes well-defined and the 
probability of the soft photon 
emission is derived from Eq.(\ref{prob2}) by integrating over the
photon energy from the small photon mass to the maximum energy 
$\omega_{max}$ of emitted soft photons

\begin{equation}
\Gamma_{soft}=\Gamma_{el} \cdot \frac{2\alpha}{\pi}ln\frac {m_{\mu}}{m_e}ln\frac{\omega_{max}}{\lambda}.
\label{soft}
\end{equation} 

\section*{Probability of radiative muon - electron conversion}

The inner bremsstrahlung spectrum diverges for low energy photons but this 
divergence in the total rate is canceled by virtual photon corrections. 
Therefore physically observable is the sum of probabilities of the elastic 
process corrected by the virtual photon emission and of the soft 
photon emission.

By adding the probabilities $\Gamma_{virt}$ and $\Gamma_{soft}$
the infrared divergences from the soft bremsstrahlung and from the
lepton vertex corrections cancel each other producing a finite 
probability of radiative $\mu \to e (\gamma)$ conversion with the
emission of soft photon with the energy below $\omega_{max}$:

\begin{equation}
\Gamma = \Gamma_{el} \cdot \left [1-\frac{2\alpha}{\pi}ln\frac {m_{\mu}}{m_e}ln\frac{m_{\mu}}{\omega_{max}} \right ].
\label{vs}
\end{equation} 

Note that this is the probability of radiative $\mu \to e (\gamma)$ conversion
calculated in the first order in electromagnetic constant $\alpha$.

Table ~\ref{table:sigcor} shows the ratio of probabilities  
$\Gamma / \Gamma_{el}$
of $\mu \to e $ conversion calculated with and without
radiative corrections. Radiative corrections include the 
virtual photon emission and soft photon emission with energy
below $\omega_{max}$.

\begin{table}[htb!]
\begin{center}
\begin{tabular}{|c|c|c|c|c|c|c|}
\hline
& & & & & &\\
$\omega_{max}$ (MeV)   &0.1 &0.5 &1 &1.5 &2 &2.5 \\
& & & & & &\\
\hline
& & & & & &\\
$\Gamma/\Gamma_{el}$    &0.828 &0.867 &0.885 &0.895 &0.902 &0.907 \\
& & & & & &\\
\hline
\end{tabular}
\caption {The ratio of probabilities  
of $\mu \to e $ conversion calculated with and without
radiative corrections versus maximum soft photon energy $\omega_{max}$ .}
\label{table:sigcor}
\end{center}
\end{table}

One can see that radiative corrections reduce the probability calculated
without radiative corrections. For $\omega_{max} = 1.5 MeV$ this
corrections is about 11$\%$.

\section*{Spectrum of conversion electrons}

In this section we consider the process of soft photon emission with
energy below $\omega_{max} = 1.5 MeV$ because as it was discussed above 
the selection criterion
of $\mu \to e (\gamma)$ conversion events is an electron momentum above
the threshold momentum of 103.5 MeV/c , which corresponds to the maximum 
energy loss of 1.5 MeV.

It is important to note that one does not need to simulate the
photon spectrum down to very small photon energies. 
A straggling in a target  leads to
a smearing of the electron momentum.
To get the electron momentum distribution from a target
a pattern recognition and track reconstruction procedure \cite{dk_seeds}
 based on the Kalman filter
technique \cite{kalman}  was applied taking into account backgrounds,
delta-rays and straw inefficiency.
It will be shown below that an electron momentum distribution from a target
can be fitted by a Gaussian 
with a standard deviation $\sigma $ = 200 keV/c.
Therefore a photon emission below 200 keV is 
indistinguishable from the elastic process of muon - electron
conversion and it is not 
required to simulate this photon emission. In the following we
will simulate photon emission in the range 
from $\omega_{min} = 100 keV$ to $\omega_{max} = 1.5 MeV$.
The particular choice of 
$\omega_{min}$ is not important because as we mentioned above
for photon energies below 200 keV
the process of $\mu \to e (\gamma)$ conversion
will look as the elastic process. Moreover a formula for the total
probability of $\mu \to e (\gamma)$ process contains logarithm of
$\omega_{min}$ and therefore it is not very sensitive to $\omega_{min}$.   

The relative probability $\Gamma_{soft} / \Gamma_{el}$ of 
photon emission in the range from $\omega_{min}$ to $\omega_{max}$ 
with respect to the probability of the corresponding elastic process
can be calculated
by integrating the photon spectrum Eq.(\ref{prob2})
of radiative $\mu \to e $ conversion over photon energy:

\begin{equation}
\frac{\Gamma_{soft}}{\Gamma_{el}} = \frac{2\alpha}{\pi} \cdot 
ln(\frac{m_{\mu}}{m_e}) \cdot ln(\frac{\omega_{max}}{\omega_{min}}) ~.
\label{prob_tot} 
\end{equation} 

\begin{figure}[htb!]
  \centering
  \includegraphics[width=0.7\textwidth]{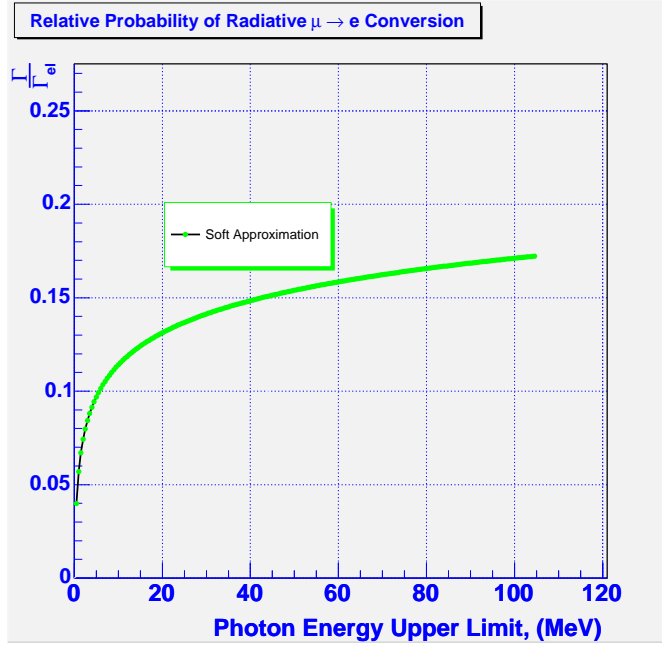}
  \caption{
Relative probability to emit a photon  
in the range from $\omega_{min}$ to $\omega_{max}$
for radiative muon-electron conversion
in the approximation of soft photon emission. 
Minimal photon energy is $\omega_{min} = 100 ~keV$.
 }
\label{fig:wtot_mue}
\end{figure}

Figure ~\ref{fig:wtot_mue} shows the relative probability to
emit a photon for the radiative $\mu \to e $ conversion.

Table ~\ref{table:wtot} shows the relative probability  
to emit a photon  
in the range from $\omega_{min}$ to $\omega_{max}$
for radiative muon-electron conversion calculated by Eq.(\ref{prob_tot}).

\begin{table}[htb!]
\begin{center}
\begin{tabular}{|c|c|c|c|c|c|c|c|}
\hline
& & & & & & &\\
$\omega_{max}$ (MeV)   &0.5 &1 &1.5 &2 &3 &4 &5 \\
& & & & & & &\\
\hline
& & & & & & &\\
probability ($\%$)    &4 &5.7 &6.7 &7.4 &8.4 &9.1 &9.7 \\
& & & & & & &\\
\hline
\end{tabular}
\caption { Relative probability to emit a photon  
in the range from 100 keV to $\omega_{max}$ .}
\label{table:wtot}
\end{center}
\end{table}

Let's consider two subranges
of the photon energy: below 100 keV and from 100 keV to 1.5 MeV.
Below 100 keV we consider the process as the elastic one 
occurring according to Table  ~\ref{table:sigcor} for
$\omega_{max}$ = 100 keV with the relative rate 
$\Gamma / \Gamma_{el}$ = 82.8$\%$.  

Photons emitted in the subrange 100 keV - 1.5 Mev
carry away a part of energy released in $\mu \to e (\gamma)$ conversion. This
leads to the appearance of low energy tail in the electron distribution.
According to Table ~\ref{table:wtot} with the relative rate  
$\Gamma_{soft} / \Gamma_{el}$ = 6.7$\%$
in the process of $\mu \to e (\gamma)$ conversion the electron energy is
distributed in the range 103.5 - 104.863 MeV.

In Monte Carlo simulations the elastic process is simulated
with the probability $82.8/(82.8+6.7)\% \approx 92.5\%$ and the process of soft photon emission
in the range from 100 keV to 1.5 MeV with the probability $6.7/(82.8+6.7)\% \approx 7.5\%$. 
The total number of useful $\mu \to e (\gamma)$ conversion events 
including the elastic process and one photon emission is 
$N = (0.828 + 0.067)N_0 = 0.895N_0$ where 
$N_0$ is the number of elastic events calculated without radiative 
corrections.
We take into account the photon energy distribution  in the  simulation by
sampling the photon spectrum (\ref{prob2}) according to the equation
$\omega = \omega_{min} \cdot (\frac{\omega_{max}}{\omega_{min}})^r$
where r is distributed uniformly from 0 to 1.

\begin{figure}[htb!]
  \centering{\hbox{
  \includegraphics[width=0.5\textwidth]{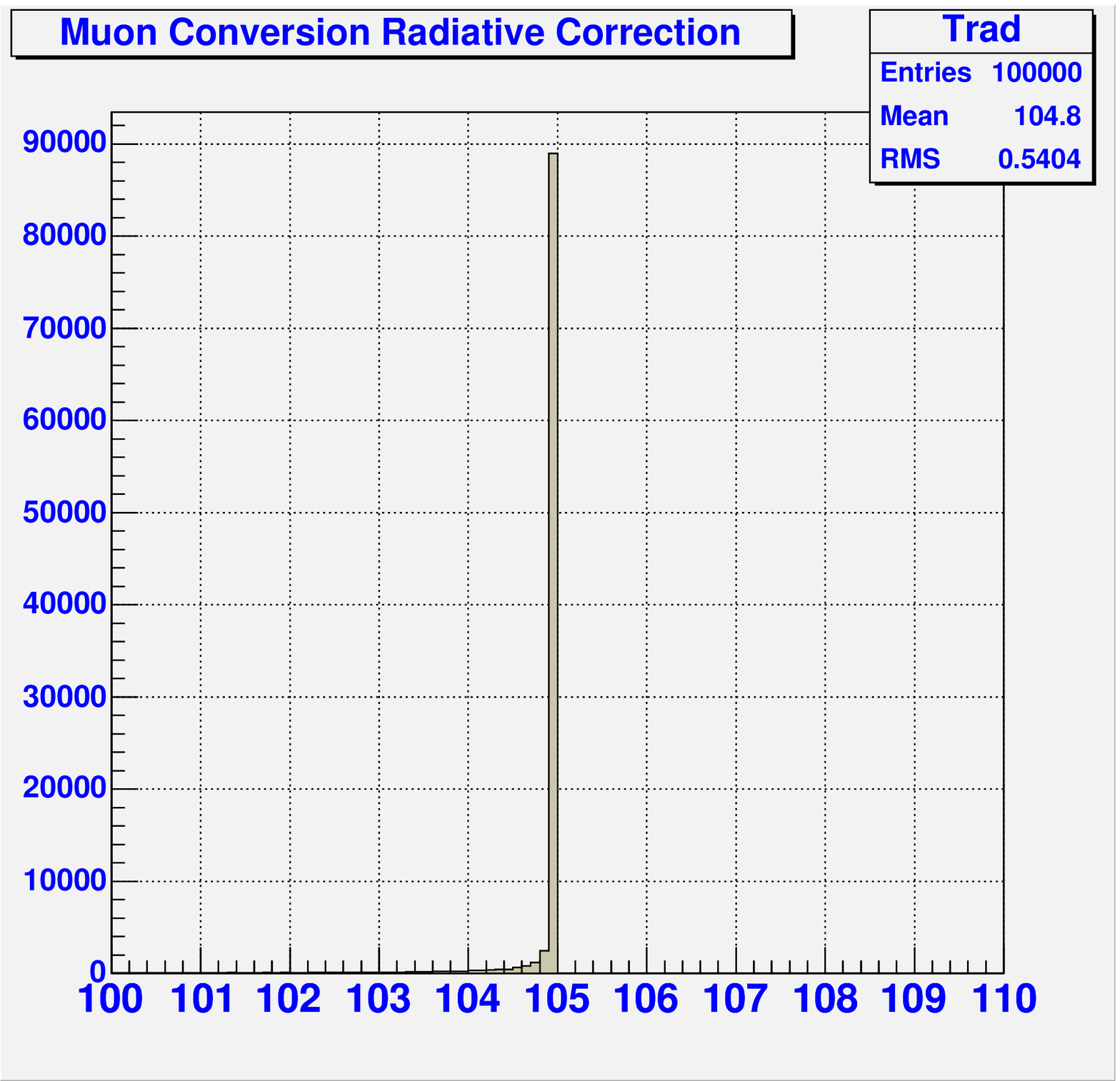}
  \includegraphics[width=0.5\textwidth]{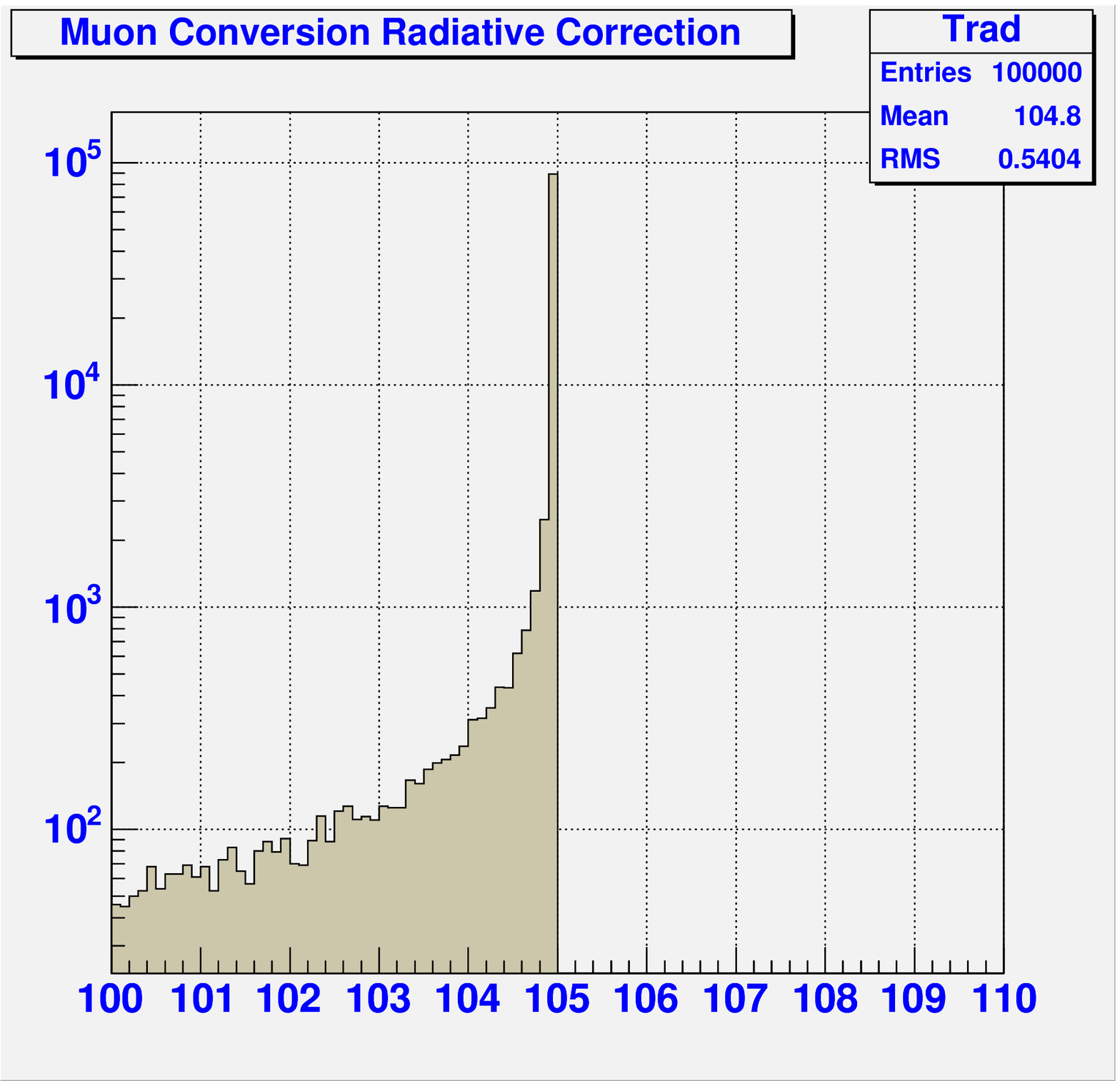} }}
  \caption{
Momentum distribution of electrons produced  in
the process of $\mu \to e (\gamma)$ conversion in linear and
logarithmic scale. Soft photon emission is taking into account.
 }
\label{fig:p_mue}
\end{figure}

The spectrum of electrons produced in the process
of radiative $\mu \to e $ conversion is shown in Figure ~\ref{fig:p_mue}
in linear (left) and logarithmic (right) scale.
Due to the photon emission a low energy tail in the electron spectrum 
appears.

\begin{figure}[htb!]
  \centering{\hbox{
  \includegraphics[width=0.5\textwidth]{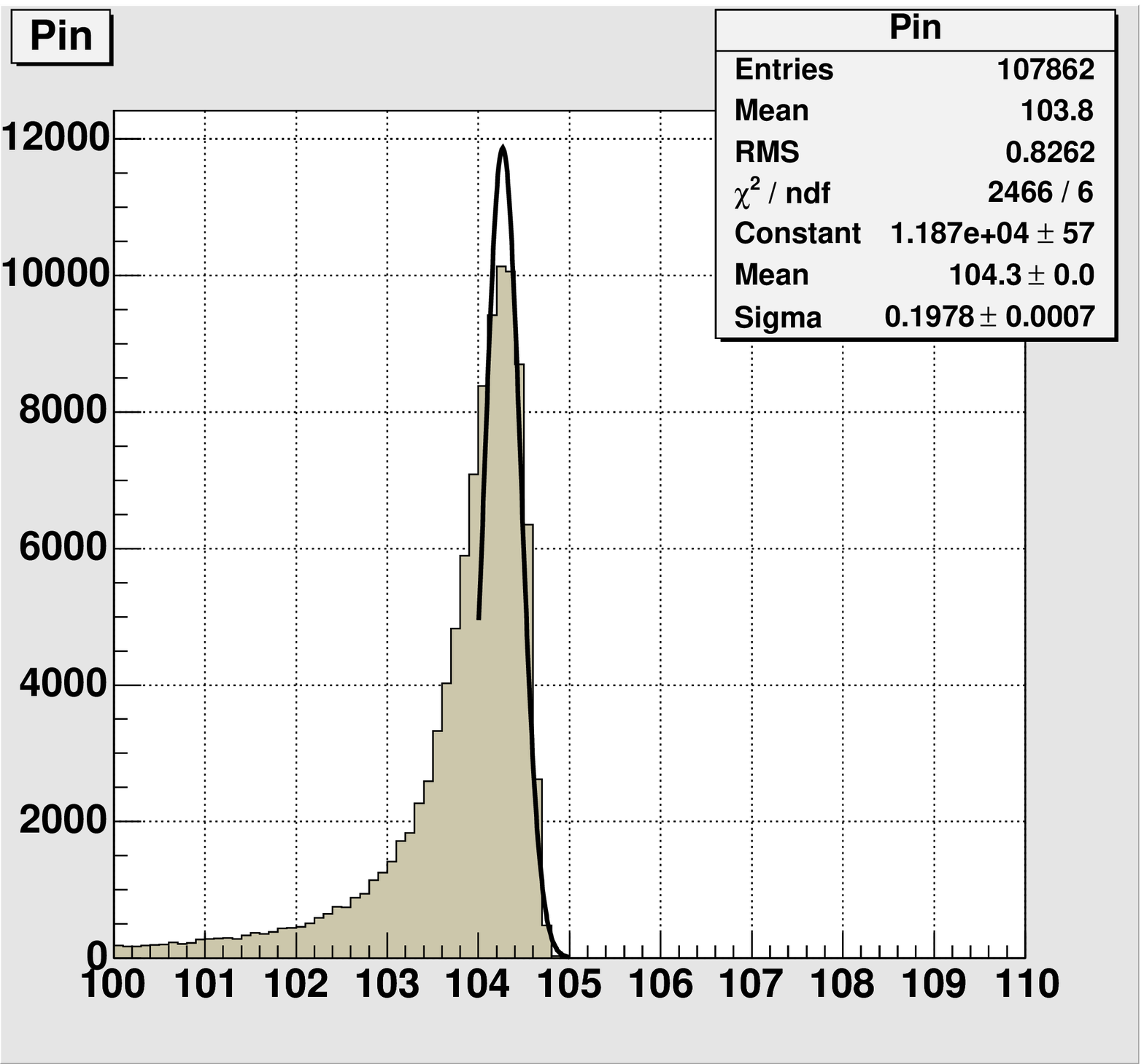}
  \includegraphics[width=0.5\textwidth]{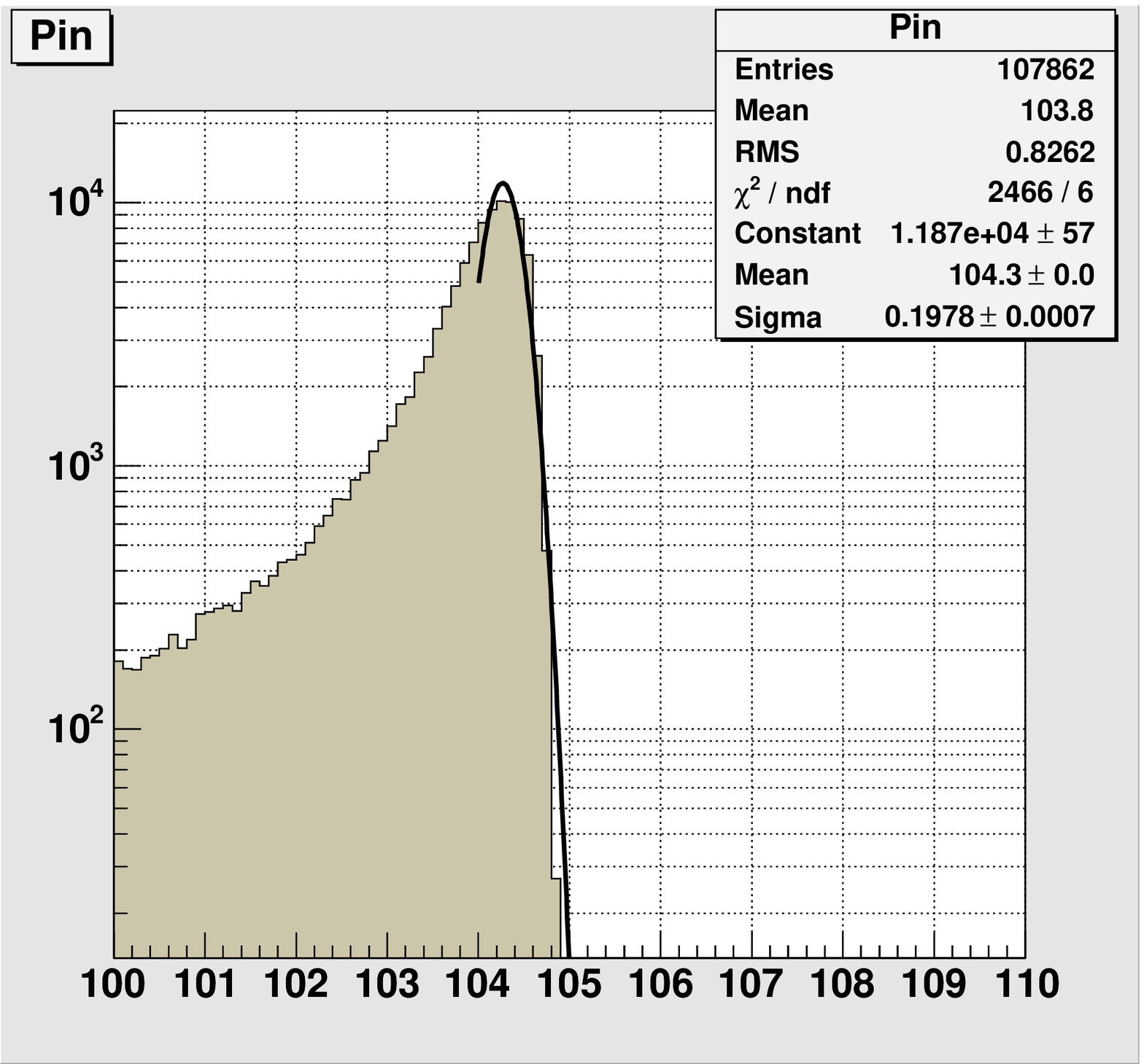} }}
 \caption{
 Momentum distribution of electrons from a target. This 
distribution does not include a photon emission.
 }
\label{fig:Pin}
\end{figure}

The momentum distribution of electrons from a target
smeared by straggling
is shown in Figure ~\ref{fig:Pin} in linear (left) and logarithmic (right) scale
by neglecting the photon emission. 

By comparing of Figures ~\ref{fig:p_mue} and ~\ref{fig:Pin}
one can see that the effect of the photon emission is small
in comparison with the effects of straggling in the target.

In the soft photon approximation the average energy of photons
emitted in the range from $\omega_{min}$ to $\omega_{max}$

\begin{equation}
\bar \omega = \Gamma^{-1} \int_{\omega_{min}}^{\omega_{max}} \omega \frac{d\Gamma_{soft}}{d\omega} d\omega =
\frac {\frac{2\alpha}{\pi}ln\frac {m_{\mu}}{m_e} \cdot (\omega_{max}-\omega_{min})}
{1-\frac{2\alpha}{\pi}ln\frac {m_{\mu}}{m_e}ln\frac{m_{\mu}}{\omega_{max}}} .
\end{equation}

For $ \omega_{min} \ll \omega_{max}$ = 1.5 MeV this equation 
gives $\bar \omega \approx$ 0.04 MeV.

The classical formula from ~\cite{jackson} also describes 
correctly the soft photon emission
but the extension of this formula to high energy photons
is a rough approximation.
This is especially important in
calculations of the average radiated energy because this quantity
depends crucially on the high energy behavior of the spectrum.
In order to describe the high energy end of the photon spectrum
specific particle physics models are required. 

\section*{Inner bremsstrahlung in the radiative pion decay}

In order to understand qualitatively a hard photon emission and to validate
the range of application of soft photon approximation
the radiative pion decay is considered.
In previous section the spectrum of soft photons was found for the 
process of $\mu \to e (\gamma)$ conversion. In order to get predictions for the high energy
part of the photon spectrum one has to relay on a variety of specific models for
the $\mu - e$ vertex. In order to avoid complications in analysis of the high energy
photon spectrum we just consider the inner bremsstrahlung in
the  process of radiative pion decay 
that possesses similar kinematic
behavior and similar mass scale as $\mu \to e (\gamma)$ conversion.
Equations describing the photon and electron spectra for the radiative 
pion decay are given in Appendix.   

Figure ~\ref{fig:dwdx} shows the differential photon spectrum 
in relative photon energy $x = 2\omega /m_{\pi}$ 
for the inner bremsstrahlung in 
the radiative pion decay. Lower curve is the result of 
calculations based on the exact formula of Appendix. Upper curve is
obtained in the approximation of soft photon emission where only a term 
proportional to $x^{-1}ln(m_{\pi}/m_e)$ contributes.

\begin{figure}[htb!]
  \centering
  \includegraphics[width=0.7\textwidth]{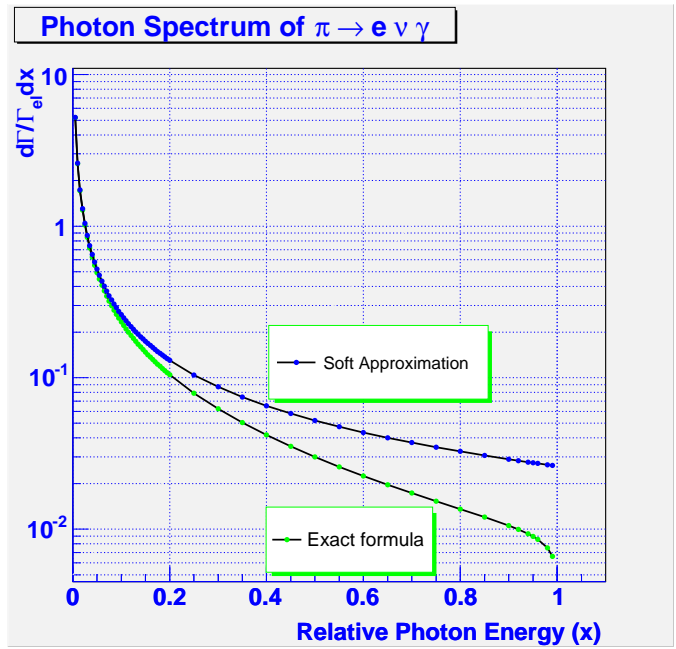}
  \caption{
Differential spectrum in relative photon energy $x = 2\omega /m_{\pi}$ 
for the inner bremsstrahlung in 
the radiative pion decay. Lower curve is based on the exact formula,
upper curve is obtained in the approximation of soft photon emission. 
 }
\label{fig:dwdx}
\end{figure}

It follows from this plot that 
up to the photon energy $\omega$ = 5 MeV the exact photon 
spectrum can be approximated
by soft photon emission $1/\omega$ with the precision better than $10\%$. 
At high energies the spectra differ significantly because as x tends to the 
kinematic boundary $x = 1 - \Delta^2$ where $\Delta = m_e/m_{\pi}$ 
the exact spectrum tends to 0. 

This behavior of the photon spectrum is of general nature. In particular 
in the case of $\mu \to e (\gamma)$ conversion low
energy spectrum is exactly the same if one uses the proper mass scale. 
At high energies one could expect qualitatively similar 
photon spectra for $\mu \to e (\gamma)$ conversion and radiative pion decay.

\begin{figure}[htb!]
  \centering
  \includegraphics[width=0.7\textwidth]{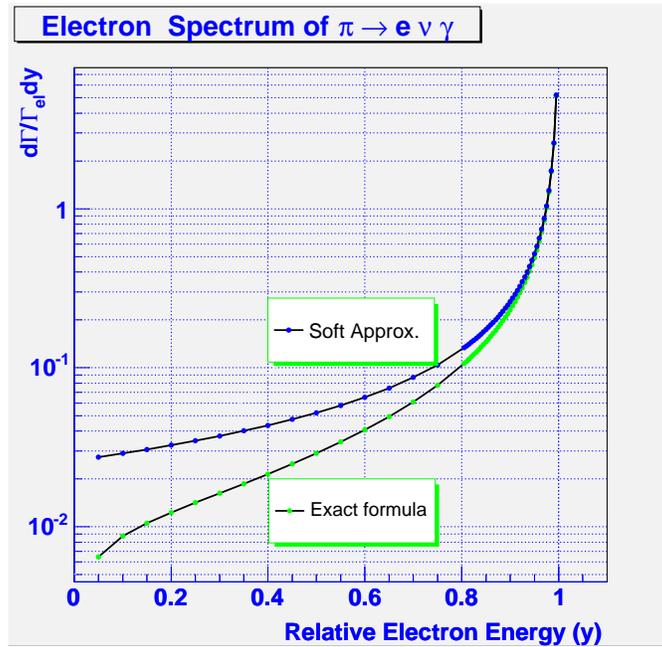}
  \caption{
Differential spectrum in relative electron energy $y = 2E /m_{\pi}$ 
for the inner bremsstrahlung in 
the radiative pion decay. Lower curve is based on the exact formula,
upper curve is obtained in the approximation of soft photon emission. 
 }
\label{fig:dwdy}
\end{figure}

Figure ~\ref{fig:dwdy} shows the differential electron spectrum 
in relative electron energy $y = 2E /m_{\pi}$ 
for the inner bremsstrahlung in 
the radiative pion decay. Lower curve is based on the exact formula
of Appendix,
upper curve is obtained in the approximation of soft photon emission. 
This spectrum is divergent as y tends to $1 + \Delta^2$ because 
this limit corresponds to the emission of low energy photons.

By integrating the photon spectrum of radiative pion decay over photon
energy one gets a relative probability $\Gamma / \Gamma_{el}$ of 
photon emission in the range from $\omega_{min}$ to $\omega_{max}$ 
with respect to the probability of the corresponding elastic process
(see Appendix).
Figure ~\ref{fig:wtot_pi} shows the relative probability of photon 
emission for the radiative
pion decay. Minimal photon energy is $\omega_{min} = 100 ~keV$.
Lower curve is based on the exact formula,
upper curve is obtained in the approximation of soft photon emission. 

\begin{figure}[htb!]
  \centering
  \includegraphics[width=0.7\textwidth]{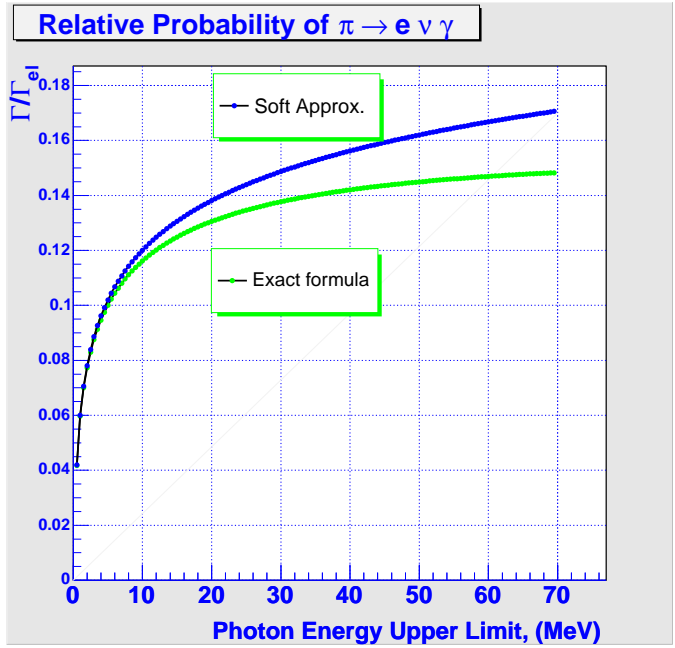}
  \caption{
Relative probability to emit a photon  
in the range from $\omega_{min}$ to $\omega_{max}$ for the radiative
pion decay. Minimal photon energy is $\omega_{min} = 100 ~keV$.
Lower curve is based on the exact formula,
upper curve is obtained in the approximation of soft photon emission. 
 }
\label{fig:wtot_pi}
\end{figure}

In particular, it follows from Figure ~\ref{fig:wtot_pi} that the 
relative probability to produce a photon in the 
energy range 100 keV - 1.5 MeV is
7$\%$ which is comparable, due to an increase 
in a rate of photon emission at low energies,
with the probability of 7.8$\%$ to produce a
high energy photon in much broader range 1.5 MeV - 70 MeV.
Note that the range 100 keV - 1.5 MeV is described with high
precision (better than 1$\%$) by the approximate formula
of soft photon emission.

A comparison of this plot with the 
corresponding Figure ~\ref{fig:wtot_mue} for $\mu \to e (\gamma)$ 
conversion,
which was obtained in the approximation of soft photon emission by
Eq.(\ref{prob_tot}) shows that the relative
probabilities $\frac{\Gamma}{\Gamma_{el}}$ of these processes are
quite close even though the energies released in these processes 
are somewhat different. 

\section*{Conclusion}

This study has shown that
radiative corrections including a virtual photon correction and 
soft photon emission 
below 1.5 MeV lead to a reduction by about 10$\%$ in
the probability of $\mu \to e $ conversion process calculated 
without radiative corrections. It is not necessary to consider
a photon emission above 1.5 MeV because these events do not
survive a cut in the electron momentum of 103.5 MeV.  

The soft photons emission below 1.5 MeV
contributes to a change of electron spectrum from monoenergetic one 
at 105 MeV to a spectrum with a low energy tail. 
The effect of smearing of the initial momentum distribution
due to the soft photon emission is small in
comparison with a smearing due to energy loss fluctuations in the target. 
The average energy of photons emitted
below 1.5 MeV is found to be small of about 40 keV.

By analyzing the radiative pion decay which possesses the same kinematic 
properties and a comparable energy scale as $\mu \to e (\gamma)$ conversion it was found that 
up to the photon energy $\omega$ = 5 MeV the exact photon spectrum can be approximated
by soft photon emission $dN/d\omega \sim 1/\omega$ with the precision better than $10\%$. 

We wish to thank A.Mincer and P.Nemethy
for fruitful discussions and helpful remarks.

\newpage
\section*{Appendix  \\ Spectra of the radiative pion decay}

According to ~\cite{brown} the double differential rate of 
pion decay including the inner bremsstrahlung with respect
to photon energy $\omega$ and electron energy E is given by

\begin{equation}
\frac{d^2\Gamma}{dx dy} = \frac{\alpha}{2\pi} \cdot \frac
{\Gamma(\pi \to e \nu)}{1-\Delta^2} \cdot I_{IB}
\label{double_diff}
\end{equation}.

where $\Gamma(\pi \to e \nu)$
is the decay rate for non-radiative pion decay, 
$\Delta=m_e /m_{\pi}$.

In Eq.(~\ref{double_diff}) $I_{IB}$ is

\begin{equation} 
I_{IB}=\left [ \frac{1-y+\Delta^2}{x^2(x+y-1-\Delta^2)} \right ]
\cdot \left [ x^2+2(1-x)(1-\Delta^2)+\frac{2x \Delta^2 
(1-\Delta^2)}{x+y-1-\Delta^2} \right ]
\label{I_IB}
\end{equation}

where  
$x = 2\omega /m_{\pi}$ and $y = 2E /m_{\pi}$ are 
relative photon and electron energies, respectively. These energies 
are directly measurable in radiative pion decay.

Differential distributions in photon energy $\omega$ 
and electron energy E are obtained from Eq.(~\ref{double_diff})
by integrating over corresponding kinematic regions.

It follows from the conservation of four-momenta $p_{\pi}=k+q+p_e$,
where $p_{\pi}, k, q, p_e$ are the four-momenta of 
pion, neutrino, photon, and electron, respectively, that

\begin{center}
$k^2 = (p_{\pi}-q-p_e)^2 = m_{\pi}^2-2m_{\pi}(\omega+E)+
2\omega (E-|p_e|z)+m_e^2 = 0$
\end{center}

where $z=cos\theta$ , $\theta$ is the angle between the electron
and photon momenta.

Since $|z| \le 1$ the previous condition implies that

\begin{center}
$(1+\Delta^2-x-y+xy/2)^2-x^2(y^2/4-\Delta^2) \le 0$.
\end{center}

The ranges of definition of x and y are easily established from
this condition:

\begin{center}
$2\Delta \le y \le 1+\Delta^2 , 
1-y/2-(\sqrt (y^2-4\Delta^2))/2 \le x \le 1-y/2+(\sqrt (y^2-4\Delta^2))/2$
\end{center}

or

\begin{center}
$0 \le x \le 1-\Delta^2 ,
(1+\Delta^2-2x+x^2)/(1-x) \le y \le 1+\Delta^2 .$
\end{center}

Additional conditions were used to get ranges above:
$y^2-4\Delta^2 \ge 0$, $(p_{\pi}-p_e)^2 \ge 0$, 
$(1+\Delta^2-2x+x^2)/(1-x) \le 1+\Delta^2$.

By integrating the double differential rate Eq.(~\ref{double_diff}) in 
corresponding limits one gets for the photon spectrum

\begin{eqnarray}
\nonumber \frac{d\Gamma}{dx} = \frac{\alpha}{2\pi} \cdot \frac
{\Gamma(\pi \to e \nu)}{1-\Delta^2} \cdot ~~~~~~~~~~~~~~~~~~~~~~~~\\
\nonumber \left [
(x-2(1-\Delta^2)+2(1-\Delta^2)^2/x)ln(\frac {1-x}{\Delta^2})- 
x(1-\Delta^2/(1-x)) \right ]
\end{eqnarray}

and for the electron spectrum

\begin{eqnarray}
\nonumber
\frac{d\Gamma}{dy} = \frac{\alpha}{2\pi} \cdot \frac
{\Gamma(\pi \to e \nu)}{1-\Delta^2} \cdot \frac {1}{1+\Delta^2-y} \cdot (A + B)
\end{eqnarray}
\begin{eqnarray}
\nonumber
A = 2(1-\Delta^2)(2\Delta^2-y)ln \left [\frac 
{(1-y/2+\sqrt(y^2-4\Delta^2)/2)^2}{1+\Delta^2-y} \right ] 
\end{eqnarray}
\begin{eqnarray}
\nonumber 
B = (1+\Delta^2(-2+5\Delta^2)-4\Delta^2y+y^2)
ln \left [\frac 
{(y/2-\Delta^2+\sqrt(y^2-4\Delta^2)/2)^2}{\Delta^2(1+\Delta^2-y)} \right ].
\end{eqnarray}

\end{document}